\documentclass[11pt, a4paper]{amsart}

\usepackage{amsmath}
\usepackage{latexsym}
\usepackage[all]{xy}
\usepackage{color}
\newtheorem{teorema}{Theorem}[section]
\newtheorem{definicion}[teorema]{Definition}
\include{amslatex}
\newtheorem{proposicion}[teorema]{Proposition}
\newtheorem{lema}[teorema]{Lemma}

\newtheorem{comentario}[teorema]{Remark}

\begin{document}
\title[On the notion of Randers space and gauge invariance]{On the interplay between the notions of Randers space and gauge invariance}

\maketitle
\begin{center}
\author{Ricardo Gallego Torrom\'e\footnote{email: rigato39@gmail.com}}
\end{center}
\begin{center}
\address{Department of Mathematics\\
Faculty of Mathematics, Natural Sciences and Information Technologies\\
University of Primorska, Koper, Slovenia}
\end{center}

\begin{abstract}
 In this paper we argue that when gauge invariance is taken into consideration, there is no consistent geometric framework of Finsler class that can accommodate Randers type spaces. In this context, an alternative non-Finslerian framework for Randers spacetimes compatible with gauge invariance is introduced.
\end{abstract}

\section{Introduction}
G. Randers \cite{Rand} introduced an asymmetric theory of spacetime, encoding the asymmetry of time evolution. Since them, Randers spaces have become a paradigm of non-Riemannian/Lorentzian Finsler spaces.
However, issues related with the positivity and non-degeneracy
of the resultant metric, specially the discussion of the gauge invariance and its physical implications, remain open.

This situation is totally unsatisfactory, since it is through its relation with $U(1)$-electrodynamics by which Randers spacetimes partially acquires special relevance in theoretical physics.
The problem is analyzed in this paper through considering the problem of degeneracy induced by gauge transformation in the fundamental tensor. The problem is considered in detail in two standard paradigms of Finsler spacetimes, Asanov's \cite{Asa} and Beem's \cite{Asa,Beem1} theories, but also briefly considering other frameworks \cite{Javaloyes Sanchez,Lamerzahl Perlick Hasse}. Our analysis strongly suggests that gauge transformations makes Randers spaces incompatible with Finsler frameworks.

Since there are situations where gauge symmetry is significant, but also one would like to have a geometric description of formal Lorentz force equation as a geodesic equation, a new mathematical formulation for Randers spaces, whose critical points correspond to the solutions of the Lorentz force equation, is of relevance.
 As an attempt to investigate this problem, we consider in {\it section 3} a refinement of the usual Randers-type space which is compatible with the gauge invariance. However, our theory imply that a Randers structure is essentially non-metric, since they will appear singularities in the domain of definition of the fundamental tensor.

\section{Lorentzian Randers space as Finsler spacetime structure}
For Randers-type metrics with Lorentzian signature, the possible degeneracy of the fundamental tensor associated with a Randers function is a problematic issue. The situation is further complicated
 when Randers-type geometries are applied to model the electrodynamics of point charged particles
 interacted with external fields, since then gauge transformations can make the fundamental tensor degenerate. In this {\it section} we discuss these issues and we argue why it is not suitable to consider Randers spaces as Finsler spacetimes from this point of view.

In this section we consider Randers spaces from the optics of several proposed Finsler spacetime frameworks.
Special attention has paid to Asanov's formalism \cite{Asa} and
Beem's formalism \cite{Beem1} theories, but the reader will observe that the same conclusions extend to other frameworks proposed in the literature \cite{Azami Javaloyes,Caponio Stancarone,Javaloyes Sanchez, Lamerzahl Perlick Hasse,Minguzzi2017, Minguzzi2019, Hohmann Pfeifer Voicu, Pfeifer 2019}. For Randers type spaces these formalisms are not completely
satisfactory. Asanov's formalism does not allow for intrinsically defined lightlike curves. A main objection to Beem's formalism is that there is not known a natural definition of
Randers-type metric in such framework. However, the main point discussed in this section concerns with the compatibility of gauge transformation and Finslerian frameworks of geometry for Randers space. This has to do with the regularity condition of Hessian tensors, a condition that appear in these approaches to Finslerian theories. Such conditions are violated by the action of gauge transformations.

\subsection{Mathematical formalisms for Finsler spacetime structures}
Let $M$ be a $\mathcal{C}^{\infty}$ smooth $n$-dimensional manifold and
TM its tangent bundle with projection $\pi:TM \to M$.  We will consider specific sub-bundles $\tilde{N}$ of the tangent bundle $TM$ whose fibers over $x\in \, M$ is $\tilde{N}_x$. Given a local coordinate system $(U,x)$ on M,
the induced {\it natural coordinate system} on TM is $(T_{U}M,(x^0,...,x^{n-1},y^0,...,y^{n-1}))$.
In local representations, a tangent vector $y$ at the
point $x\in\,M$ is represented as the derivation $y=\sum^{n-1}_{i=0}\,y^i\frac{\partial}{\partial x^i}\big|_x$ and its natural coordinates on the natural coordinate chart $(TU, (x,y))$ as an element of $TM$ are $(x^0,...,x^{n-1},y^0,...,y^{n-1})$.

Let us consider first the following two standard definitions of
semi-Finsler structures currently being used in the literature:

\begin{definicion}Asanov's definition \cite{Asa}.
Let $\hat{\pi}_{\tilde N}:\tilde N\to M$ be a sub-bundle of TM.
A semi-Finsler structure $F$ defined on the $n$-dimensional manifold
$M$ is a function  $F:\tilde N\to
]-\infty, 0]$ such that:
\begin{enumerate}
\item It is smooth on $\tilde N\setminus\{0\}$,

\item It is positive homogeneous of degree $1$ in $y$,
\begin{align*}
F(x,{\lambda}y)=\lambda F(x,y),\,\,\,\forall \lambda
>0,
\end{align*}
\item The vertical Hessian matrix
 \begin{align}
g_{i j}(x,y): =\frac{1}{2}\frac{{\partial}^2 F^2
(x,y)}{{\partial}y^{i} {\partial}y^{j} }
\end{align}
is non-degenerate over $\tilde N\setminus\{0\}$.
\end{enumerate}
\label{definicionasanov}
\end{definicion}
$g_{ij}(x,y)$ is the matrix of the {\it fundamental tensor} $g$ and plays an analogous role to the metric tensor in Lorentzian geometry.

Vectors $y\in\tilde{N}_x\setminus\{0_x\}$ are called {\it timelike vectors} at $x$; a lightlike vector is such that $F(x,y)=0$. In Asanov's theory, there is not a well defined notion of spacelike. Furthermore, there is no notion of lightlike smooth curve, since the regularity conditions of $F$ and $g$ are restricted to $\tilde{N}\setminus\{0\}$.

\begin{definicion} Beem's definition \cite{Beem1}.
A semi-Finsler structure defined on the $n$-dimensional manifold
$M$ is a continuous function  $L:TM\to \mathbb{R}$
such that

\begin{enumerate}
\item It is $\mathcal{C}^{\infty}$ in the slit tangent bundle $\hat{N}:=
TM\setminus \{0\}$

\item It is positive homogeneous of degree $2$ in $y$,
\begin{align*}
L(x,{\lambda}y)=\lambda^2\, L(x,y),\,\,\,\forall \lambda >0,
\end{align*}
\item The Hessian matrix
 \begin{align}
g_{ij}(x,y):=\,\frac{1}{2}\frac{{\partial}^2 L (x,y)}{{\partial}y^{i}
{\partial}y^{j} }
\end{align}
is non-degenerate on $\hat{N}$.
\end{enumerate}
\label{definicionbeem}
\end{definicion}
We can call a function $L$ according to the above definition a {\it Beem's function}. Similarly as in Asanov's theory, $g$ is the fundamental tensor.
In Beem's theory there are well defined notions of timelike, lightlike spacelike vectors and smooth curves.

\subsection{Differences between Asanov's and Beems's formalisms} As we mention above, a mayor difference in the above theories are in the admissible causal notions that they support.
In Beem's framework there is a geometric definition of lightlike vectors and it is possible to derive the
Finsler geodesic equation from a
variational principle \cite{Perlick} also to include the case of lightlike geodesics. By construction, in Asanov's formalism
it is not possible to consider lightlike curves in an intrinsic way, because
lightlike vectors are excluded from the domain of regularity of the function $F$.

Second, in Asanov's framework, given a parameterized
curve $\sigma:{\bf I}\to M,\, {\bf I}\subset \mathbb{
R}$ on the semi-Finsler manifold $(M,F)$ such that
$\dot{\sigma}\in \tilde N$ for all $t\in {\bf I}=[a,b]$, the length
functional acting on $\sigma$ is given by the line integral
\begin{align}
\sigma\mapsto \mathcal{T}_{F_A}(\sigma):=\int^{b}
_{a} F (\sigma (t),\dot{\sigma}(t)) dt.
\label{funtional Asanov}
\end{align}
Because the homogeneity condition of the Finsler function $F$,
$\mathcal{T}_{F_ A}(\sigma)$ is a re-parametrization invariant
functional. On the other hand, if we consider Beem's definition, the
energy functional is given by the following expression:
\begin{align}
\sigma \mapsto \mathcal{T}_B(\sigma):=\int^{b}
_{a} {L (\sigma(t),\dot{\sigma}(t))} dt.
\end{align}
Beem's energy functional is not
re-parametrization invariant, due to the fundamental function $L$
is positive $2$-homogeneous  in $y$-coordinates.

 A third difference between Asanov's and Beem's formalisms emerges when we consider the category of {\it Randers-type} spaces. In Asanov's framework a Lorentzian Randers space should be defined in the following terms:
\begin{definicion}(Lorentzian Randers space as Asanov's Finsler spacetime)
A Lorentzian Randers space is characterized by a
semi-Finsler function $F_R:\tilde{N}\to ]-\infty,0]$ of the form
\begin{align}
F_R(x,y)=\sqrt{-\eta(x)_{ij} y^i y^j}\,+A(x,y),
\label{Randers space Asanov}
\end{align}
where $\eta:=\,{\eta}_{ij}(x)dx^i\otimes dx^j$ is a Lorentzian metric
defined on M and $A(x,y):=A _i (x)y^i$ is the value of the
action of the $1$-form $A(x)=A_i(x) dx^i$ on the tangent vector $y^i\frac{\partial}{\partial x^i}\,\in T_x
M$.
\label{asanov-randers}
\end{definicion}
In the positive definite signature case, when $\eta$ is a Riemannian
metric, the requirement for a Randers space that $g_{ij}$ is non-degenerate
implies that the $1$-form $(A _1,...,A _n )$ is bounded by ${\eta}$ \cite{BaoChernShen},
\begin{align}
A _i A _j {{\eta}}^{ij}< 1,\,\quad \quad {\eta}^{ik}{\eta}_{kj}=\delta^i\,_j.
\label{criterioranders}
\end{align}

There are several issues with the definition \ref{asanov-randers}. First,
in the case of Lorentzian signature case there is not a natural
Riemannian metric that induces a norm. Therefore, the
  criterion \eqref{criterioranders} for non-degeneracy of Lorentzian Randers spaces is absent.

 Second, for both the positive and indefinite metric, only
   the variation of the length functional \eqref{funtional Asanov}
(not directly the integrand itself) is invariant under
 the gauge transformation $A\to A+d\lambda$; the Finsler function \eqref{Randers space Asanov}
 is not gauge invariant: even if one can define a norm and a criteria for non-degeneracy of the metric,
a transformation of the $1$-form $A$
 by a gauge transformation can change the norm and the new hessian $(g_{ij})$, making it non-degenerated at a given point $x_0$. The degeneracy of the fundamental tensor precludes the construction of any connection at $x_0$. Since this can happen for every point $x_0$, it seems unnatural to have a geodesic equation which is gauge invariant, but a Finsler function which is not gauge invariant and with the potential degeneracy issue under gauge transformations.
  This fact suggests that while the variation of the
 length functional \eqref{funtional Asanov} could have physical meaning, the fundamental function $F_R(x,y)$ does not have such category.

 On the other hand, as we mention before, in Beem's formalism there is not known definition of a Finsler spacetime whose geodesics curves are also the curves of a time like Randers space geodesic.

 Note that the spoiling of regularity of the fundamental tensor by gauge transformations raised in the second point raised above applies to both, positive definite signature and Lorentzian signature metrics, as long as the gauge transformations have associated a physical, relevant meaning.
For instance, in the positive signature case, the form can be interpreted as the Fermat metric in conformally stationary spacetimes \cite{Perlick 1990a, Perlick 1990b, Perlick, Javaloyes 2009}. In this case, arbitrary gauge transformations $A\to A+d \lambda$ will describe  different classical physical situations, even if the corresponding action functional is gauge invariant. In contrast, for positive definite metrics in three dimensions, the $1$-form $A$ can represent the vector potential of a static magnetic field $dA$, while in the case of four dimensions and in the setting of Lorentzian signature, $dA$ is the Faraday form describing an electromagnetic field. In these two cases, gauge transformations of the potential $A$ provides an equivalent description of the physical fields. Hence such transformation should be allowed and embraced by any consistent geometric description of Randers spaces.

The above discussion shows that when invariance under $A\to A+d\lambda$ is of physical relevance, then there is an intrinsic problem with degeneracies of the fundamental tensor associated with such gauge transformations. This issue applies to both, positive definite and Lorentzian signature spaces.
Although we have discuss this problem in two particular frameworks of Finsler spacetimes, namely, Asanov's theory and Beem's theory, the problem is present in  other Lorentz-Finsler frameworks. Take for instance the theory of Lorentz-Finsler spaces developed in \cite{Lamerzahl Perlick Hasse}. Their definition of Finsler space is a generalization from Beem's definition allowing for singularities of measure zero on $TM$. Again, the problem is that gauge transformations precludes the regularity of the fundamental tensor, since by gauge transformations the fundamental tensor can becomes singular almost everywhere. The general problem of modelling spaces where the gauge invariance $A\to \,A+d\lambda$ is of significance is that in the definition of the fundamental tensor (as Hessian of the Finsler function or its square), the $1$-form $A$ is involved in such a way that the Hessian of the transformed $1$-form can be degenerated almost everywhere, as we will discuss in the next section.
Similar remarks apply to conic metric frameworks \cite{Javaloyes Sanchez} and in general, to the frameworks where there is a regularity condition depending upon the $1$-form $A$. By the same reasoning, other more sophisticate theories of Finsler  spacetimes fail to accommodate such gauge transformations \cite{Minguzzi2017,Pfeifer 2019}.

In resume, gauge invariance spoils a possible geometric formulation of Randers spaces as Finsler type spaces.  However, in the cases when gauge invariance is of relevance, a geometric formulation of Randers type models is certainly of interest. The development of the rudiments of such a theory by considering fundamental notions and methods of sheaf theory \cite{Hirzebruch} is intended in the next section.

\section{Gauge invariant theory of (Lorentzian) Randers spaces}
In this {\it section} we discuss a new definition of spacetimes compatible with the structures of Randers space and develop the theory of its geodesics. We show that the theory developed is consistent with gauge invariance. We restrict our considerations to time-like curves. The further extension to lightlike curves probably requires further modifications of the framework and will not be addressed here.
\subsection{Definition of Lorentzian Randers spaces $(M,\eta,[A])$}
Let us consider a smooth Lorentzian structure $(M,\eta)$. Let $\mathcal{F}(M)$ be the algebra of real functions over $M$.
The null-cone at the point $x$ is the sub-manifold of ${T}_xM$ given by
\begin{align*}
 {NC}_x:=,\{y\in T_xM\,|\,\eta(y,y)=0\}.
\end{align*}
 The null-cone bundle over M is
\begin{align*}
NC:=\,\bigsqcup_{x\in M}\,NC_x.
\end{align*}
 The unit hyperboloid bundle at the point is the sub-manifold of $T_xM$ given by
 \begin{align*}
\Sigma_x:=\{y\in \,T_xM\,|\eta_{ij}(x)y^iy^j=1\,\}.
 \end{align*}
 The unit tangent bundle over $M$ is
 \begin{align*}
\Sigma:=\,\bigsqcup_{x\in M}\,\Sigma_x.
 \end{align*}
In addition to the semi-Riemannian structure $(M,\eta)$ we assume that
a non-zero closed $2$-form ${\bf F}$ is defined on $M$. Given this setting, one possibility to define a Randers space is
Then we can propose the following notion of Lorentzian Randers spaces,
\begin{definicion}
A Lorentzian Randers space consists of a triplet $(M,\eta ,{\bf
F})$, where $M$ is a manifold, $\eta$ is a
Lorentzian metric smooth on ${TM}\setminus\{0\}$ and
 ${\bf F}\in { \bigwedge}^2 M$ such that $d{\bf F}=0$ on $M$.
\label{firstdefinitionoflorentzianRanderspace}
\end{definicion}

The $2$-form  ${\bf F}$ is an element of the second de Rham cohomology group $H^2( M,\mathbb{R})$. Therefore,
by Poincar\'{e}'s lemma, a locally smooth $1$-form $A$
such that $dA={\bf F}$ on an open set $U$ \cite{War} is given by
\begin{align}
 A(x)=\big(\,\int^1_0\, t\,\sum^{n-1}_{k=0} x^k\,{\bf F}_{kj}(tx)\,dt\big)dx^j.
 \label{poincaresformula}
 \end{align}

Let us consider the sheaf of germs of locally smooth differential $k$-forms over M $\pi:\bigwedge^k_{loc} M\to M$ \cite{War}. For each section $\mathcal{S}\in \,\Gamma\,\bigwedge^k_{loc} M$ and $x\in\,M$ there is an open neighborhood U, such that the restriction of $\mathcal{S}$ to U coincides with a locally smooth $1$-form ${S}_{loc}$ over U.
\begin{definicion}
Two locally smooth $k$-forms ${A}_1,\,A_2\in \,\Gamma\,\bigwedge^k_{loc} M$ are equivalent iff the following two conditions hold:
\begin{enumerate}
\item $\forall \,x\in M$ there are local representations $(\,^{\mu}{A}_2,\,^\mu U_1)$ and $(\,^{\nu}{A}_2,\,^\nu U_2)$ such that $x\in\,^\mu U_1\cap \,^\nu U_2$,
\item In the intersection $^\mu U_1\cap \,^\nu U_2$, it holds that $(\,^{\mu}A)=\,(d(\,^{\mu\nu}\lambda) +\,^{\nu}A)$, for a smooth function $^{\mu\nu}\lambda$.
\end{enumerate}
\label{equivalencerelationA}
\end{definicion}
Given $A\in \,\Gamma\,\bigwedge^k_{loc} M$, $[A]$ is the equivalence class of $A$.
\begin{definicion}
A Lorentzian Randers space consists of a triplet $(M,\eta ,[A])$,
where $M$ is a spacetime manifold, $\eta$ a
Lorentzian metric smooth on $TM$ and $A\in\,\Gamma\,\bigwedge^1_{loc}M$.
\label{secondefinitionofLorentzianRanderspace}
\end{definicion}
\begin{proposicion}
Definitions \ref{firstdefinitionoflorentzianRanderspace} and \ref{secondefinitionofLorentzianRanderspace} are equivalent.
\end{proposicion}
\begin{proof} Given a triplet $(M,\eta ,[A])$, one can construct a triplet  $(M, \eta, {\bf F})$ by defining ${\bf F}=dA$. By the equivalence relation \ref{equivalencerelationA}, the $2$-form ${\bf F}$ does not depend on $A$. Conversely, given the pair  $(M, \eta, {\bf F})$, for any point $x\in M$, one can construct via Poincare's formula \eqref{poincaresformula} a locally smooth $1$-form $A_{loc,x}$, defined in a (convex) domain $U(A,x)$. Then it is clear that the collection $\{A_{loc,x},\,x\in M\}$ defines a locally smooth $1$-form $A\in \,\Gamma \,\bigwedge^1_{loc} M$ with equivalence class being $[A]$. \end{proof}

For each of the local representatives
$A\in\,[\tilde{A}]$ one can define in an open set $U\subset \,M$ the
function ${F}_{A}$ as
\begin{align}
{F}_A(x,y)=\left\{
\begin{array}{l l}
\sqrt{\eta_{ij}(x)y^i y^j}+A_i(x)y^i &\quad \textrm{for  } \,\,\eta_{ij}(x)y^iy^j > 0, \\
A_i(x)y^i & \quad \textrm{for} \,\,\eta_{ij}(x)y^iy^j=0,\\
 \sqrt{-\eta_{ij}(x)y^i y^j}+A_i(x)y^i &\quad \textrm{for  } \,\,\eta_{ij}(x)y^iy^j < 0. \\
\end{array} \right.
\label{gauge Randers space}
\end{align}
 $F_{A}$ defines a sheaf homomorphism from $\Lambda^1_{loc}M$ to $\Lambda^0_{loc}(TM\setminus NC)$.
The proof of the following proposition can be obtained easily,
\begin{proposicion}
Let $(M,\eta, [A])$ be a Lorentzian Randers space with $\eta$ a
semi-Riemannian metric,
 $A\in [A]$ and $F_A$ given by equation $(3.1)$. Then
\begin{enumerate}
\item $F_A$ is smooth in $TM\setminus NC$ and is not smooth but continuous in $NC$.

\item $TM\setminus NC$ is an open fibre subbundle of ${T M}$
 and any local representation $F_A$ is locally smooth on $TM\setminus NC$.

\item The function $F_A$ is positive homogeneous of degree $1$ in $y$.
\end{enumerate}
\end{proposicion}
 For both definitions \ref{firstdefinitionoflorentzianRanderspace} and \ref{secondefinitionofLorentzianRanderspace}, fixed a local representative $A$ of $[A]$, the associated fundamental tensor
\begin{align*}
g_{ij}(x,y)=\,\frac{1}{2}\frac{\partial^2 F^2_A(x,y)}{\partial y^i\partial y^j}
\end{align*}
to the local representation $F_A(x,y)$ could be degenerated. This degeneracy problem is related with the freedom on the selection of the local forms $A$. However, the same {\it gauge freedom} that we have on the choice of $A$ provides a way to avoid such a problem. In particular, we discuss next how to extract Euler-Lagrange equations for functionals that are integrals of $F_{A}$ along curves.

\subsection{A variational principle for Lorentzian Randers spaces}
The space of admissible curves is denoted by $\widehat{\Omega}(M,t_1,...,t_N)$ and will be the space of compact, smooth, time-like curves of $\eta$ on {
M} passing through the same points $\sigma(t_k)$ at the instants $\{t_1,.....,t_N\}$. Let us consider one curve $\sigma:[a,b]\to M$, an open cover of the image $\sigma([a,b])\subset\,\tilde U(\sigma)=\,\cup^N_{h=1} \,U_B$ and a collection of local smooth potentials $\{A_1,....,A_N\}$ defined on the open sets
$\{U_1$,...,$U_N\}$ respectively. One can denote the collection $A:=\{A_1,....,A_N\}$, a {\it local cover representation} of $[A]$ along $\sigma$ (or just a representation of $[A]$ along ${\sigma}$).
\begin{definicion}
  The {\it time functional} acting on $\sigma:[a,b]\to M$ with local representation $A=\{A_1,....,A_N\}$ is
  \begin{align}
  \mathcal{T}_{F_A}(\sigma,A_1,...,A_N):=\,\sum^N_{h=2}\,\int^{t_h} _{t_{h-1}}\,
F_{A_h}(\sigma(\tau),\dot{\sigma}(\tau))\,d\tau.
\label{finslertime}
\end{align}
\end{definicion}
\begin{proposicion}
The {\it time functional} of a curve $\sigma\in\,\widehat{\Omega}(M,t_1=a,...,t_h=b)$ has the following characteristics:
\begin{enumerate}
\item $\mathcal{T}_{F_A}(\sigma,A_1,...,A_N)$ depends on the set of local $1$-forms $\{A_1,...,A_N\}$.

\item The functional is additive: given to curves $\sigma_1:{\bf I}_1\to M\in \,\widehat{\Omega}(M,a_1=t_1,...,t_{N_1}=b_1)$ and $\sigma_2:{\bf I}_2\to M\in\,\widehat{\Omega}(M,a_2=\tilde{t}_2,...,\tilde{t}_{N_2}=b_2)$ such that $\sigma_1(b_1)=\,\sigma_2(a_2),$ and the local potential to the covers are $\{A_{11},...,A_{1N_1}\}$ and $\{A_{21},...,A_{2N_2}\}$,  then
\end{enumerate}
\begin{align}
\nonumber \mathcal{T}_{F_A}(\sigma_1\circ \sigma_2,A_{11},...,A_{1N_1},A_{21},...,A_{2N_2})= & \,\mathcal{T}_{F_A}(\sigma_1,A_{11},...,A_{1N_1})\\
 &+\,
\mathcal{T}_{F_A}(\sigma_2,A_{21},...,A_{2N_2}).
\label{aditiviteofthefermattime}
\end{align}
\end{proposicion}
\begin{proof}The first property follows directly from the definition.
 The second property follows from the additivity of the line integral.\end{proof}

The condition that guarantees that the critical points of the functional $\mathcal{T}_{F_{\tilde{A}}}$ correspond to the solutions of the Euler-Lagrange equations is that the vertical Hessian $g_{ij}(x,y)$ must be non-degenerate.
 Given a representation $A$ of $[A]$, it is not guaranteed that the Hessian of $F_A$
   is non-degenerate in the domain of definition of $A$. However, because the possibility of perform {\it gauge transformations} in
    the gauge potential $A(x)\mapsto A(x)+d\lambda(x)$ it is always possible to choose local representations such that the Euler-Lagrange equations can be extracted from a variational principle in some case.
Because the additive property of $\mathcal{T}_{F_A}$, it is only necessary to discuss the case when there are only two local representative $^{\mu}A$ and $^{\nu}A$ overlapping over an open neighborhood, where $\mu$ and $\nu$ labeling arbitrary local representatives of $[A]$.
Given two curves $\sigma_1,\sigma_2:[a,b]\to M$, let us denote the {\it variation} of the {\it time functional} functional \eqref{finslertime} to the difference
\begin{align}
\delta\,\mathcal{T}_{F_{^\mu A}} &:=\,\mathcal{T}_{F_{A}}(\sigma_1,\,^{\mu}A)-\,\mathcal{T}_{F_{A}}(\sigma_2,\,^{\mu}A).
\label{variationfinslertimefunctional12}
\end{align}
\begin{lema}
Let us consider  two representatives $\{\,^{\mu}A\}$ and $\{\,^{\nu}A\}$ of $[A]$ and  two curves $\sigma_1, \sigma_2:[a,b]\to M$ with the same initial and final points $\sigma_1(a)=\sigma_2(a)$ and $\sigma_1(b)=\sigma_2(b)$ contained in $^\mu U \cap\, ^\nu U$ If the potentials are related by a gauge function $^{\mu}{A}=\,^{\nu}A+d(\,^{\mu\nu}\lambda)$, then  $\delta\,\mathcal{T}_{F_{^\mu A}}=\delta\,\mathcal{T}_{F_{^\nu A}}$.
\label{lemasobregaugetransformation}
\end{lema}
\begin{proof}
Let us consider two curves $\sigma_1, \sigma_2:[a,b]\to M$ with the same initial and final points $\sigma_1(a)=\sigma_2(a)$ and $\sigma_1(b)=\sigma_2(b)$. Then
\begin{align*}
\delta\,\mathcal{T}_{F_{^\mu A}} &=\,\mathcal{T}_{F_{A}}(\sigma_1,\,^{\mu}A)-\,\mathcal{T}_{F_{A}}(\sigma_2,\,^{\mu}A)\\
& =\,\int_{\sigma_1}\,d{t}\,F_{^\mu A}(\sigma_1({t}),\dot{\sigma}_1(t))-\,\int_{\sigma_2}\,d{t}\,F_{^\mu A}(\sigma_2({t}),\,\dot{\sigma}_2(t))\\
& =\,\int_{\sigma_1}\,d{t}\,F_{^\nu A}(\sigma_1({t}),\dot{\sigma}_1)-\,\int_{\sigma_2}\,d{t}\,F_{ ^\nu A}(\sigma_2({t}),\sigma_2(t) )\,+\int_{\sigma_1-\sigma_2}\,dt\,\frac{d\lambda (t)}{dt}\\
& =\, \mathcal{T}_{F_{A}}(\sigma_1,\,^\nu A)-\,\mathcal{T}_{F_{A}}(\sigma_2,\,^\nu A)
+\int_{\sigma_1-\sigma_2}\,dt\,\frac{d\lambda (t)}{dt}\\
&  =\delta\,\mathcal{T}_{F_{^\nu A}},
\end{align*}
the last equality because Stokes's theorem. \end{proof}

Let us consider a compact curve $\sigma:[a,b]\to M$ with a finite open cover $\tilde U(\sigma)=\,\cup^N\,_{h=1}U_h$ of $\sigma([a,b])$ and the points $\sigma(t(h,h+1))\in\,U_h\cap\,U_{h+1}\cap\sigma([a,b])$, for each possible pair of open sets $U_h$ and $U_{h+1}$ such that the above intersection is not empty. Since $\sigma([a,b])$ is compact and the open cover finite, the number of points $\{\sigma(t(\mu,\nu))\}$ can be chosen finite.
\begin{definicion}
An allowed variation of $\sigma:[a,b]\to \,M$ with intersecting points $\{\sigma(a)=\sigma(t(0,1)),...,
\sigma(t(h,{h+1})),....,\sigma(b)=\sigma(t(N-1,N))\}$ is a smooth map
\begin{align*}
\Lambda :(-\varrho_0,\varrho_0)& \times\, [a,b] \to M
\end{align*}
such that:
\begin{enumerate}
\item Each curve $\Lambda(\varrho,\cdot)\in\,\hat{\Omega}(M, t_1=a,...,t_N=b)$,

\item The central curve is $\sigma$, $\Lambda(0,t)=\sigma(t)$,

\item Each curve $\Lambda(\varrho,\cdot)$ is such that $\Lambda(\varrho,t(\mu,\nu))=\,\sigma(t(\mu,\nu))$.
\end{enumerate}
\label{alowedvariation}
\end{definicion}
\begin{comentario}
Note that the variations that we are considering are with fixed initial and final points. However, we also require intermediate fixed points. This is an essential requirement for our proof on the gauge invariance of the variation of $\mathcal{F}_A$. Therefore, the curves of the allowed variation $\Lambda(\cdot,t)$ are on {\it zig-zag} across the central curve  $\sigma:[a,b]\to M$.
\end{comentario}

Let us consider two local representations of $[A]$ associated with the finite open cover $\{U_h\}$ of the variation $\Lambda (\rho,t)$ such that the local representations $\{A_1,...,A_N\}$ and $\{\tilde{A}_1,....,\tilde{A}_N\}$ of $[A]$ are related by gauge transformations,
\begin{align*}
A_{h}=\,\tilde{A}_{\tilde{h}}+\,d\lambda_{h\tilde{h}}.
 \end{align*}
 Also let us assume that for each open domain $U_h$ of the cover of the variation $\Lambda(\rho,t)$, each pair of local representations $(A_h,\tilde{A}_h)$ contains the point $\sigma(t(h,h+1))$ on the intersection of $U_h\cap U_{h+1}$, which is specified on the definition of the variation $\Lambda(\rho,t)$.
\begin{proposicion}
Let $({M,\eta, [A])}$ be a Lorentzian Randers space.
Then the variation of the Finslerian time functional $\mathcal{T}_{F_A}(\Lambda(\rho,t),A_1,...,A_N)$ respect to an allowed variation $\Lambda(\rho,t)$ of a curve $\sigma$ is gauge invariant: for two gauge related local representations  $\{A_1,...,A_N\}$ and $\{\tilde{A}_1,....,\tilde{A}_N\}$, the following equality holds:
\begin{align}
\delta\mathcal{T}_{F_A}(\Lambda(\rho,t),A_1,...,A_N)=\,\delta\mathcal{T}_{F_A}(\Lambda(\rho,t),\tilde{A}_1,...,\tilde{A}_N)
\end{align}
\end{proposicion}
\begin{proof}
By {\it lemma} \ref{lemasobregaugetransformation}, the property is true when restricted to each {\it sector} of the curves $\Lambda(\cdot,t)$ on each $U_h$. Then the property follows by the additivity property of $\mathcal{T}_{F_A}$.
\end{proof}
\begin{lema}
Let $(M,\eta, [A])$ be a Lorentzian Randers space and $\sigma:[a,b]\to M$ a time-like curve such that $\eta(\dot{\sigma},\dot{\sigma})=1$. Then for each point $\sigma(t)$ there exists a local representative ${A}$ of $[A]$  along $\sigma:[a,b]\to M$ such that the Hessian of the
 function $F^2_{{A}}$ is non-degenerate.
\label{lemasobrelecciondelgauge}
\end{lema}
\begin{proof} By {\it lemma} \ref{lemasobregaugetransformation}, $\mathcal{T}_{F{A}}$ depends only on
the selection of the $1$-forms $^{\mu}A$ along the image $\sigma([a,b])$. Let us assume first that we only require one local form to cover the whole chart. If the Hessian of $F^2_{A}$ is non-degenerate, then the thesis is satisfied.
The requitement that the Hessian of $F^2_{{A}}$
     is non-degenerate is equivalent to the following mathematical condition:
\begin{align*}
\phi(x,y,\lambda=0):=\,2+\frac{{A}_i(x)y^i+\,\sqrt{|\eta_{ij}(x)y^iy^j|}\,(\eta^{ij}\,{A}_i(x)
 {A}_j(x))}{\sqrt{|\eta_{ij}(x)y^iy^j|}+\,{A}_i(x)y^i}\,>0.
\end{align*}
This is identical to the condition for positive definite Randers spaces \cite{BaoChernShen,Hohmann Pfeifer Voicu}.

Instead, let us assume that the Hessian for $F^2_A$ is degenerated.
 Let us consider a gauge transformation of the form
\begin{align*}
\tilde{A}_i=A_i(x)\,+\partial_i \lambda(x)).
\end{align*}
The expression to be considered is now of the form
\begin{align}
\phi(x,y,\lambda)=\,2+\frac{\tilde{A}_i(x)y^i+\,\sqrt{|\eta_{ij}(x)y^iy^j|}\,(\eta^{ij}\,\tilde{A}_i(x)
 \tilde{A}_j(x))}{\sqrt{|\eta_{ij}(x)y^iy^j|}+\,\tilde{A}_i(x)y^i}.
\label{expresionforepsilon}
\end{align}
If $F^2_{\tilde{A}}$ is degenerated, the necessary and sufficient condition for the Hessian $F^2_{\tilde{A}}$ not to be degenerated can be expressed as
\begin{align}
\phi(x,y,\lambda=0)\,\phi(x,y,\lambda)<\,0.
\label{condition degeneration non degeneration}
\end{align}
Let us consider $\lambda=\,\epsilon^{-1}\,\tilde{\lambda}$ with $\epsilon>0$ constant and the corresponding asymptotic expansion in terms of $\epsilon\to 0$ in the expression \eqref{condition degeneration non degeneration}. It is easy to observe that the asymptotic expansion $\epsilon \to 0$ implies the following relation:
\begin{align*}
&\phi(x,y,\lambda=0)\,\phi(x,y,\lambda) =\,\left(\frac{1}{1+\frac{1}{\epsilon}\,\frac{\tilde{\chi}}{\beta(d\tilde{\lambda})}}\right)\,\cdot\\
&\cdot\left(\frac{1}{\epsilon}\,\chi_{-1}(x,y)\beta(\tilde{\lambda})\,\eta^{-1}(d\tilde{\lambda},d\tilde{\lambda})
 +\chi_0(x,y)+\,\epsilon \,\chi_1(x,y)\left(\beta(d\tilde{\lambda})\right)^{-1}\,\right) ,
\end{align*}
 where $\chi_{-1},\,\chi_0(x,y),\,\chi_1(x,y)$ are positive homogeneous functions of degree zero in $y$ the variable and $\tilde{\chi}$ is positive homogeneous of degree zero. Because such homogeneity, the above expression lives in the compact sphere bundle $SM$, which is compact. Therefore, $\chi_{-1},\,\chi_0(x,y),\,\chi_1(x,y)$ are bounded functions. In the asymptotic limit $\epsilon\to 0$,
  \begin{align}
  \phi(x,y,\lambda=0)\,\phi(x,y,\lambda) \xrightarrow{\epsilon \to 0}\,|\eta(y,y)|^{1/2}\,\eta^{-1}(\tilde{\lambda},\tilde{\lambda})\,\chi(x,y,A),
  \end{align}
  where
  \begin{align*}
  \chi(x,y,A)=\,2\,|\eta(y,y)|^{1/2}+\,3\,y^iA_i+\,|\eta(y,y)|^{1/2}\,\eta^{-1}(A,A).
  \end{align*}
Since $\chi$ does not depend upon the $1$-form $\tilde{\lambda}$ and since $\tilde{\lambda}$ is an arbitrary smooth function on an open set of $M$, one can always choose $\tilde{\lambda}$ such that the product $\phi(x,y,\lambda=0)\,\phi(x,y,\lambda)$ is negative.

The above arguments applies to each  local representative $A$  of the class [A].
By compactness of $\sigma([a,b])$, it is only necessary to consider a finite number of local representatives $\{A_1,...,A_h\}$, for which the above argument is applied to each of the representatives individually.
\end{proof}
\begin{comentario}
Let us remark that the methodology of the proof of lemma \ref{lemasobrelecciondelgauge} also implies  that starting with a $1$-form $A$ such that the Hessian $F^2_{A}$ is non-degenerated, one could potentially reach degeneration by changing the gauge.
\end{comentario}
Finally, we show that the theory developed here offers a geometric consistent description of Randers spaces with gauge symmetry. Specifically, the equation of motion can be re-written as a critical condition in a integral functional,
\begin{teorema}
Let $(M,\eta, [A])$ be a Lorentzian Randers space and the local representations $\{A_1,...,A_N\}$ such that the Hessian of $F^2_{{A}}$ is not degenerate on each of the open sets $\{U\}^N_{h=1}$. Then the critical curves of the functional $\mathcal{T}_{F_{A}}$ are well defined.
\label{teoremasobreEulerlagrangelorentz}
\end{teorema}
\begin{proof}
Let us consider a local representation along $\sigma:[a,b]\to M$ as in {\it lemma} \ref{lemasobrelecciondelgauge} that requires two local $1$-forms. Since the Hessian matrix of $\mathcal{F}_{A}$ is non-degenerate in the respective domains, the variation of the functional time $\mathcal{F}_A$ with respect to an allowed variation and the condition of criticality implies that
\begin{enumerate}
\item On $U_{\mu}$, the Euler-Lagrange equation holds for $F_{{A}}(\sigma\cap\,U_\mu,\,^\mu A))$, except maybe on the point $\sigma(t(\mu,\nu))$.
 \item On $U_{\nu}$, the Euler-Lagrange equation holds for $F_{{A}}(\sigma\cap\,U_{\nu},\,^\nu A)$, except maybe on the point $\sigma(t(\mu,\nu))$.
\end{enumerate}
Then by continuity, for a critical curve, the Euler-Lagrange equations must also hold on the point $\sigma(t(\mu,\nu))$.
Given a local representation $A$ the Euler-Lagrange equation for $F_A$ corresponds to the Lorentz force equation for $F=dA$ is a gauge invariant equation, both equations coincide on the overlap, since on the overlap, the gauge potential are related by a gauge transformation, $^\mu A=\,^\nu A+\,d(^{\mu \nu}\lambda).$
 Therefore, the critical points of the functional $\mathcal{T}_{F_A}$ are well defined and it satisfies the Lorentz force equation.
\end{proof}
Because $A$ can be chosen such the Hessian of $F^2_A$ is non-degenerated, the critical condition for the functional $\mathcal{T}_{F_{A}}$ can be re-written as an Euler-Lagrange equation. But Euler-Lagrange equations can be understood as integral conditions for a geodesic spray associated with a non-linear connection \cite{MironHrimiucShimadaSabau:2002}.

 Therefore, the theory of Lorentzian Randers spaces encompasses both, gauge transformations and a geometric theory associated to the corresponding Euler-lagrange equations. This is achieved at the expenses of abandoning  a purely Finslerian formulation of the type described in Finsler spacetimes theories.

\section{Discussion}
 In this work we have examined the interplay of gauge invariance and Randers spaces as Finsler structures. Our analyse shows that gauge invariance and a Finsler type formulation of such theories are incompatible, for both, positive definite and for Lorentzian signature spaces. In the case of Lorentzian structure, our argument has been mainly reduced to investigate two theories of Finsler spacetimes, namely, Asanov's theory and Beem's theory. Although our treatment cannot be considered complete, since other theories of Finsler spacetimes have been developed, for instance \cite{Azami Javaloyes,Caponio Stancarone,Javaloyes Sanchez, Lamerzahl Perlick Hasse,Minguzzi2017, Minguzzi2019, Hohmann Pfeifer Voicu}, our heuristic argument in section 2 shows the difficulties in accommodating both gauge invariance and Randers spaces in the context of a Finslerian framework, an argument that is independent of the particular framework. Even in the positive case, where there is a satisfactory framework for Finsler spaces, it is not affordable to accommodate Randers spaces and gauge invariance, as consequence of the argument consequence of lemma \ref{lemasobrelecciondelgauge} in discussed in section 3.

Let us remark that there are situations where the gauge invariance under transformations $A\to A+d\lambda$ is  not of physical significance. This is the case, for instance, when  Randers spaces describe the Fermat metric \cite{Perlick 1990a}. In this case, a change of $A$ implies a change of physical associated system. Other examples can be found in explicit Lorentz breaking Randers type models as discussed in \cite{Chang Li}. However,
in situations when gauge invariance is of relevance and if geometric structures corresponding to the spacetime dynamics of point particles are of relevance too, then we have shown that it is necessary a non-Finslerian theory of Randers spaces and Randers spacetimes. The first obvious case is a classical point particle interacting with a classical electromagnetic potential, when no back-reaction effects are considered. In this case, the equation of motion of point test particles is the Lorentz force equation, which is gauge invariant. The other relevant case is the original suggestion of G. Randers \cite{Rand}. Randers' theory is an attempt to model non-symmetric time evolution as a local effect of the spacetime geometry. In Randers' theory, the observables able to detect non-reversibility are the time functional and the corresponding time test particle. Both correspond to gauge invariant geometric objects. However, gauge invariance spoils the corresponding geometric theory of geodesics of Randers theory as a metric space in terms of connections, as we have discussed in this paper. We think that these two examples justify the developed of a compatible theory of geometric connection compatibility with gauge transformations. Such a theory was developed in section 3.

Finally, let us remark that a natural general framework for gauge theory is sheaf theory \cite{Bott and Tu, Hirzebruch}. Actually, that sheaf theory plays a fundamental role in gauge theory is remarkable, although well known, because the theory of electromagnetic potential forms corresponds to the theory of local $1$-forms local defined on open sets over M constitute the natural example of sheaf \cite{Bott and Tu}.
  Therefore, it is not surprise that sheaf theory appears also in  formalizations of a geometric theory compatible with local gauge transformations.

\end{document}